\documentclass[11pt,a4paper]{article}

\usepackage{jcappub,amsfonts,amsmath,amssymb,bm,graphicx}

\usepackage[active]{srcltx}
\usepackage{color}

\def\10{$SO(10)$}
\def\21{SU(2) $\otimes$ U(1) }

\def\422{$SU(4) \otimes SU(2) \otimes SU(2)$}
\def\321{SU(3) $\otimes$ SU(2) $\otimes$ U(1)}

\newcommand {\ignore}[1]{}

\def\lsim{\raise0.3ex\hbox{$\;<$\kern-0.75em\raise-1.1ex\hbox{$\sim\;$}}}
\def\gsim{\raise0.3ex\hbox{$\;>$\kern-0.75em\raise-1.1ex\hbox{$\sim\;$}}}

\newcommand{\AddrAHEP}{%
  AHEP Group, Institut de F\'{\i}sica Corpuscular --
  C.S.I.C./Universitat de Val{\`e}ncia \\
  Edificio Institutos de Paterna, Apt 22085, E--46071 Valencia, Spain}

\title{Lepton asymmetries and primordial hypermagnetic helicity
  evolution }
\author[a,b] {V. B. Semikoz}
\author[c] {D.D. Sokoloff}
\author[a] {J.~W.~F.~Valle}
 
\affiliation[a]{\AddrAHEP}
\affiliation[b]{ IZMIRAN, Troitsk, Moscow region,
  142190, Russia}
\affiliation[c]{Department of Physics,
  Moscow State University, 119999, Moscow, Russia}
\emailAdd{semikoz@ific.uv.es}
\emailAdd{sokoloff@dds.srcc.msu.su}
\emailAdd{valle@ific.uv.es}

\abstract{The hypermagnetic helicity density at the electroweak phase
  transition (EWPT) exceeds many orders of magnitude the galactic
  magnetic helicity density. Together with previous magnetic helicity
  evolution calculations after the EWPT and hypermagnetic helicity
  conversion to the magnetic one at the EWPT, the present calculation
  completes the description of the evolution of this important
  topological feature of cosmological magnetic fields. It suggests
  that if the magnetic field seeding the galactic dynamo has a
  primordial origin, it should be substantially helical. This should
  be taken into account in scenarios of galactic magnetic field
  evolution with a cosmological seed.}


\begin{document}

\maketitle

\section{Introduction}

The magnetic helicity is a relatively new and very attractive point of
interest in cosmic magnetohydrodynamics (MHD) and dynamo theory. The
point is that the magnetic helicity $H=\int{\bf AB}{\rm d}^3x$ where
${\bf B}$ is the magnetic field and ${\bf A}$ is the vector potential
is an integral of motion in MHD in the absence of viscosity (inviscid
case). It looks natural to base an understanding of cosmic MHD on
balance equations for conserved quantities such as magnetic field
energy and magnetic helicity. It has been noted that the magnetic
helicity conservation is much more restrictive in astrophysical
objects than the energy conservation
~\cite{Shukurov:2005pf,Brandenburg:2006fg}.  The point is that usually
there is a huge supply of kinetic energy in the form of a general
rotation of a celestial body and it is quite easy to imagine a
spectral energy flux which influences energy balance including
magnetic field energy.

The nature of the initial fields (and corresponding magnetic
helicities) that seed subsequent dynamo or turbulent amplifications is
largely unknown~\cite{Kulsrud:2007an,Kronberg:1993vk}. It might be
that the seed fields are produced during epoch of galaxy formation
from frozen-in magnetic fields of protogalaxy experiencing
gravitational collapse, or ejected by the first supernovae or active
galactic nuclei. Let us call this as the standard astrophysical
scenario A. Alternatively the seed fields might originate from much
earlier epochs of the Universe expansion, down to the cosmological
inflation phase transition epoch~\cite{Grasso:2000wj}. Let us call
that as the cosmological scenario B.  The standard view in scenario A
is that the magnetic field evolution in the early Universe (or an
astrophysical object) starts from a state with almost vanishing
magnetic helicity. The dynamo amplification of a seed field in
astrophysical scenarios produces, however, a large-scale magnetic
field with substantial magnetic helicity. In such case one must
compensate it providing a contribution of magnetic helicity of
small-scale magnetic fields. Then one faces with a severe problem of
how to redistribute magnetic helicity over the desired scales in order
to keep it small. Moreover, it is difficult to prevent a disastrous
dynamo suppression of large-scale magnetic helicity by helical
small-scale magnetic field. Alternatively in scenario B, if the seed
magnetic field contained a lot of magnetic helicity one can use that
as the helicity required to obtain the desired large-scale field at
later epochs. This makes the dynamo generation of galactic magnetic
fields much less constrained than in the standard scenario A. If the
galactic dynamo exploits the primordial magnetic helicity then one
must expect that a large-scale galactic magnetic field has also a
preferable sign of helicity. Observations show indeed some hint that
one of the possible helicity signs seems preferable
\cite{krause-beck}.

Relying on scenario B with the hypermagnetic field evolution passing
through the electroweak phase transition (EWPT) epoch, we explore here
the magnetic helicity generation in the early universe.  This suggests
as a new alternative the possibility that the magnetic field starts
from a state with a substantial supply of primordial magnetic
helicity. Indeed, a very small fluctuation of a seed hypermagnetic
field at very early epochs before the EWPT woul be exponentially
amplified due to the presence of a large right electron asymmetry,
$\xi_{eR}=\mu_{eR}/T\neq
0$~\cite{Giovannini:1997eg,Semikoz:2007ti,Semikoz:2011tm} and hence
acquire a huge initial magnetic helicity by the EWPT epoch and
subsequently. This is possible only for hypermagnetic fields having a
non-trivial topological structure with non-vanishing linkage number
$n\neq 0$ ~\cite{Akhmet'ev:2010ba}. The hypermagnetic helicity density
given by the product $h_Y={\bf Y}\cdot{\bf B}_Y$, hence proportional
to a large hypermagnetic field squared value $h_Y\sim
B^2_Y\Lambda$. This gets transformed to the magnetic helicity, which
can therefore be much larger than the helicity density associated to
galactic magnetic fields, $h_{Y}\gg h_{\rm gal}$ where $h_{\rm
  gal}\sim B^2_{\rm gal}l_{\rm gal}$. For $B_{\rm gal}\simeq
10^{-6}~G$, $l_{\rm gal}=100~kpc=3\times 10^{23}~cm$, this is
estimated as $h_{\rm gal}=3\times 10^{11}~G^2$cm. Thus, the primordial
magnetic helicity can be the main supply for magnetic helicity of
galaxies.  The goal of this paper is to provide a complete description
of magnetic helicity evolution passing through the EWPT epoch (we also
comment on later epochs within the causal picture).  In Section 2 we
calculate the hypermagnetic helicity in the symmetric phase of the hot
primordial plasma, using the corresponding solution of the Faraday
equation. In Section 3 we review the evolution of the hypermagnetic
helicity passing through the EWPT and in Section 4 we make the final
comments on our results and on the perspectives for the cosmological
origin of helical galactic magnetic fields.

\section{Hypermagnetic helicity}
\label{sec:hyperm-helic}

In the comoving frame ${\bf V}=0$ \footnote{Note that common expansion
  can be easily taken into account via conformal coordinates with the
  change of the cosmological time $t$ to the conformal one $dt\to
  a(t)d\eta$.} the Faraday induction equation governing the evolution
of hypermagnetic fields ${\bf B}_Y=\nabla\times {\bf Y}$ reads
\begin{equation}\label{Faraday}
  \frac{\partial {\bf B}_Y}{\partial t}=\nabla\times \alpha_Y{\bf B}_Y + \eta_Y\nabla^2{\bf B}_Y,
\end{equation}
where the hypermagnetic helicity coefficient
$\alpha_Y$~\cite{Semikoz:2011tm} is given by the right electron
chemical potential $\mu_{eR}$ and the hot plasma conductivity
$\sigma_{\rm cond}(T) \sim T$ as
\begin{equation}\label{alpha}
\alpha_Y(T)=-\frac{g^{'2}\mu_{eR}}{4\pi^2\sigma_{cond}}~~,
\end{equation}
and $\eta_Y =(\sigma_{\rm cond})^{-1}$ is the hypermagnetic diffusion
coefficient, $g^{'}=e/\cos\theta_W$ is the Standard Model U(1) gauge
coupling.  We assume that a certain right electron asymmetry $\sim
\mu_{eR}(t_0)$ at a very early cosmological epoch , $t_0\ll t_{EW}$,
has been generated by an unspecified mechanism.  This is our starting
point. Then in the presence of the hypercharge field $Y_{\mu}$ this
asymmetry $n_{eR}- n_{\bar{e}R}=\mu_{eR}T^2/6$ evolves due to the
Abelian anomaly for right electrons,
\begin{equation}\label{Abelian}
\partial_{\mu}j_{eR}^{\mu}=-\frac{g^{'2}y_R^2}{64\pi^2}Y_{\mu\nu}\tilde{Y}^{\mu\nu}, ~~y_R=-2,
\end{equation}
which evolves together with the hypermagnetic field in
Eq.(\ref{Faraday}) in a self-consistent way. Note that such coupled
evolution of $B_Y(t)$ and $\mu_{eR}(t)$ has been recently considered
in Ref. \cite{Dvornikov:2011ey} for the particular case of the
Chern-Simons wave hypermagnetic field configuration, but without
considering such important feature as the hypermagnetic helicity which
we discuss here.

Our second assumption is the presence of a non-zero initial
hypermagnetic field $B_0^Y\neq 0$. This should be a mean field with a
small amplitude provided by some stochastic distribution of
hypermagnetic fields.

The key parameter in the master Eq. (\ref{Faraday}) is the helicity
parameter for the hypermagnetic field, given in
eq.~(\ref{alpha}). This can be obtained from the Chern-Simons term in
the effective Standard Model Lagrangian density for the hypercharged
field $Y_\mu$ :
\begin{equation}\label{CS}
{\cal L}_{CS}= - \frac{g^{'2}\mu_{eR}}{4\pi^2}{\bf B}_Y{\bf Y}.
\end{equation}
An recent interpretation of the Chern-Simons anomaly parameter
$\alpha_Y$ as a polarization effect has been given in
Ref.~\cite{Semikoz:2011tm} using standard statistical averaging of the
right electron pseudovector current $<\bar{e}\gamma_j\gamma_5e>$ in the
Standard Model Lagrangian in vacuum (alternative one-loop level
calculations in finite temperature field theory were used in
\cite{Laine:1999zi,Redlich:1984md}).

Multiplying Eq.~(\ref{Faraday}) by the corresponding vector potential
and adding the analogous expression obtained from the evolution
equation governing the vector potential (multiplied by hypermagnetic
field) and integrating over space, one gets the evolution equation for
the hypermagnetic helicity ${\rm H}_Y=\int d^3x {\bf Y}\cdot{\bf
  B}_Y$ as
\begin{eqnarray}\label{helicity}
&&\frac{{\rm dH}_Y}{{\rm dt}}=
-2\int_V({\bf E}_Y\cdot{\bf B}_Y)d^3x -\oint [Y_0{\bf B}_Y + \nonumber\\
&& +{\bf E}_Y\times {\bf Y}]d^2S= -2\eta_Y (t)\int d^3x(\nabla\times {\bf B}_Y)\cdot {\bf B}_Y + \nonumber\\
&&+2\alpha_Y (t)\int d^3xB_Y^2(t).
\end{eqnarray}
Note that we have omitted in the last equality the surface integral
$\oint(...)$ since fields vanish at infinity during the symmetric
phase. However, such surface integral can be important at the
boundaries of different phases at the electroweak phase transition, $T
\sim T_{EW}$. In Ref.~\cite{Akhmet'ev:2010ba} the authors have studied
how the hypermagnetic helicity flux penetrates the surface separating
the symmetric and broken phases, and how the hypermagnetic helicity
density $h_Y={\bf B}_Y{\bf Y}$ converts into the magnetic helicity
density $h={\bf B}{\bf A}$ at the EWPT time, see also discussion in
Sec.~\ref{sec:magn-helic-evol} below.

Notice also that the evolution equation in eq.~(\ref{helicity}) is
similar to eq.~(7) in Ref.~\cite{Semikoz:2004rr}, which holds after
the electroweak phase transition, $T\ll T_{EW}$. There the point-like
short-range Fermi neutrino-plasma interaction mediated by heavy
$W,Z$-bosons was used, instead of the long-range interaction through
the massless hypercharge field $Y_{\mu}$ in the unbroken phase.

Now using the simplest solution of the Faraday equation
(\ref{Faraday}) in the $\alpha^2$-dynamo \cite{1983flma....3.....Z}
that corresponds to maximum hypermagnetic field amplification
rate~\footnote{This is the case for the particular hypermagnetic
field scale $\Lambda=k^{-1}=\kappa\eta_Y/\alpha_Y$ where
$\kappa=2$, $k$ being the Fourier wave number in ${\bf B}_Y({\bf
  x},t)= \int (d^3k/(2\pi)^3){\bf B}_Y({\bf k},t)e^{i{\bf kx}}$.},
\begin{equation}\label{extremumcase}
B_Y(t)=B_0^Y\exp \left[\int_{t_0}^t\frac{\alpha^2_Y
(t^{'})}{4\eta_Y (t^{'})}{\rm
d}t^{'}\right],
\end{equation}
we obtain from Eq. (\ref{helicity}) the hypermagnetic helicity :
\begin{eqnarray}\label{result}
&&H_Y=H_Y(t_0)+ 2(B_0^{Y})^2\int {\rm d}^3x\int_{t_0}^t{\rm
d}t^{'}\alpha_Y (t^{'})\times\nonumber\\&&\times
\exp\left[{2\int_{t_0}^{t^{'}}\left[\frac{\alpha^2_Y(t^{''})}{4\eta
(t^{''})}\right]{\rm d}t^{''}}\right],
\end{eqnarray}
where $H_Y(t_0)$ is the initial helicity value at the moment $t_0$ and
we have omitted the hypermagnetic diffusion term. The latter is
usually neglected for an ideal Maxwellian plasma $\sigma_{\rm cond}\to
\infty$, $\eta_Y\to 0$, while here we must compare in
Eq.~(\ref{helicity}) the second helicity generation term $\alpha_Y\sim
(\sigma_{\rm cond})^{-1}$ given by Eq. (\ref{alpha}) with the first
diffusion term $\eta_Y\sim (\sigma_{\rm cond})^{-1}$. Notice that in
Ref.~\cite{Semikoz:2004rr} one could neglect the diffusion term for
the Maxwellian plasma in the broken phase, $T\ll T_{EW}$, since for
that case the magnetic helicity coefficent $\alpha$
\cite{Semikoz:2003qt,Semikoz:2004rr} did not depend on the
conductivity.

The direct estimate of the relative magnitude of the terms in the
r.h.s. of Eq.  (\ref{helicity}) allows us to neglect the diffusion
term only for hypermagnetic field inhomogeneity scales obeying the
inequality:
\begin{equation}\label{scale}
\Lambda\gg \frac{\eta_Y}{\alpha_Y}~.
\end{equation}

For an arbitrary (large) scale in our causal scenario,
$l_H>\Lambda=\kappa \eta_Y/\alpha_Y\gg \eta_Y/\alpha_Y$, the
amplification of hypermagnetic field is given by \cite{Semikoz:2011tm}
\begin{eqnarray}\label{arbitrary}
&&B_Y(t)=B_0^Y\exp \left[\left(\frac{1}{\kappa} - \frac{1}{\kappa^2}\right)\int_{t_0}^{t}\frac{\alpha_Y^2(t^{'})}{\eta_Y(t^{'})}dt^{'}\right]=\nonumber\\
=&&B_0^Y\exp \left[83\left(\frac{1}{\kappa} - \frac{1}{\kappa^2}\right)\int_x^{x_0}\frac{{\rm d}x^{'}}{x^{'2}}
\left(\frac{\xi_{eR}(x^{'})}{0.0001}\right)^2\right],\nonumber\\
\end{eqnarray}
and that of the hypermagnetic helicity is given by
\begin{eqnarray}\label{arbitrary2}
&&H_Y(t)=H_Y(t_0)+ 2(B_0^Y)^2\int {\rm d}^3x\int_{t_0}^t{\rm d}t^{'}\alpha_Y(t^{'})\times\nonumber\\&&\times
\exp\left[{\left(\frac{2}{\kappa} - \frac{2}{\kappa^2}\right)\int_{t_0}^{t^{'}}\left(\frac{\alpha^2_Y((t^{''})}
{\eta(t^{''})}\right){\rm d}t^{''}}\right]=\nonumber\\&&=H_Y(t_0)+2(B_0^Y)^2\int {\rm d}^3x\int_{t_0}^t{\rm d}t^{'}
\alpha_Y(t^{'})\times\nonumber\\&&\times \exp\left[{166\left(\frac{1}{\kappa} - \frac{1}{\kappa^2}\right)
\int_{x^{'}}^{x_0}\frac{{\rm d}x^{''}}{x^{''2}}\left(\frac{\xi_{eR}(x^{''})}{0.0001}\right)^2}\right].\nonumber\\
\end{eqnarray}
Here the ratio $x=T/T_{EW}=(t_{EW}/t)^{1/2}$ is given by the Friedman law and $\xi_{eR}=\mu_{eR}/T$ is the dimensionless right electron asymmetry.
Thus, from Eq.~(\ref{arbitrary}) for the extremum value $\kappa=2$ we
obtain the strongest amplification eq.~(\ref{extremumcase}) and from
Eq. (\ref{arbitrary2}) one finds the corresponding value of the
hypermagnetic helicity in eq.~(\ref{result}).

Let us comment on our reference value choice $\xi_{eR}\sim 10^{-4}$ used in eqs. (\ref{arbitrary}),(\ref{arbitrary2}). Taking into account
the right electron (positron) asymmetry $n_{eR}-
n_{\bar{e}R}=\mu_{eR}T^2/6$ in the presence of chirality flip
processes with the rate $\Gamma$, and substituting the hyperelectric field ${\bf E}_Y= - {\bf V}\times {\bf B}_Y + \eta_Y\nabla\times {\bf B}_Y - \alpha_Y{\bf B}_Y$ ~\cite{Semikoz:2011tm} into the Abelian anomaly
Eq.~(\ref{Abelian}) rewritten in uniform medium
as $\partial_t(n_{eR} - n_{\bar{e}R})=-(g^{'2}/4\pi^2){\bf E}_Y{\bf B}_Y$
one finds the kinetic equation for $\mu_{eR}$:
\begin{equation}\label{right}
  \frac{\partial \mu_{eR}}{\partial t} = -
  \frac{6g^{'2}(\nabla\times \mathbf{B}_Y)\cdot\mathbf{B}_Y}{4\pi^2T^2\sigma_\mathrm{cond}} -
  (\Gamma_B + \Gamma)\mu_{eR}.
\end{equation}
Here the rate $\Gamma_B=6(g^{'2}/4\pi^2)^2B^2_Y/T^2\sigma_{cond}$
coming from the helicity term $\sim \alpha_Y$ occurs in strong
hypermagnetic fields much bigger than the chirality flip rate
$\Gamma$, $\Gamma_B\gg \Gamma$ ~\cite{Semikoz:2011tm}.

Under the assumption of slowly changing hypermagnetic fields
$B_Y(t)\approx const$, and choosing the Chern-Simons wave
configuration of the hypercharge field as, $$Y_0=Y_z=0,~~ Y_x=Y(t)\sin
k_0z,~~ Y_y=Y(t)\cos k_0z,$$ for which $(\nabla\times {\bf B}_Y)\cdot
{\bf B}_Y=B_Y^2(t)k_0$, $B_Y(t)=k_0Y(t)$, we can easily solve the
kinetic equation (\ref{right}) getting:
 \begin{eqnarray}\label{solution}
&& \xi_{eR}(t)=\left[\xi_{eR}(t_0) - \frac{Q}{\Gamma_B + \Gamma}\right]e^{- (\Gamma_B + \Gamma)(t-t_0)} +\nonumber\\&&+ \frac{Q}{\Gamma_B + \Gamma}\approx \frac{Q}{\Gamma_B + \Gamma}\approx \frac{Q}{\Gamma_B}= -\frac{4\pi^2k_0}{Tg^{'2}}.
 \end{eqnarray}
 Here we used notation
 $Q=-(6g^{'2}/4\pi^2T^3\sigma_{cond})(\nabla\times {\bf B}_Y)\cdot
 {\bf B}_Y=-(6g^{'2}/4\pi^2T^3\sigma_{cond})B_Y^2k_0$. In obtaining
 (\ref{solution}) we neglected the time dependence for times $t\to
 t_{EW}$ for which $\Gamma_Bt_{EW}\gg 1$. On the other hand, retaining
 the time term for the zero initial asymmetry $\xi_{eR}(t_0)=0$ we get
 from (\ref{solution}) the asymptotical growth of $\mid\xi_{eR}\mid$
 in a strong hypermagnetic field due to the Abelian anomaly:
 $$
 \mid\xi_{eR}(t)\mid=\frac{\mid Q\mid}{\Gamma_B + \Gamma}\left[1 - e^{-(\Gamma_B + \Gamma)(t - t_0)}\right]\approx$$$$\approx \frac{4\pi^2k_0}{Tg^{'2}}\left[1 - e^{-\Gamma_B(t - t_0)}\right].
 $$

 Taking into account for the survival condition of the Chern-Simons
 wave versus ohmic diffusion, $k_0<10^{-7}T$, substituting weak
 coupling $g^{'2}=0.12$ we get the estimate of the lepton asymmetry in
 a strong hypermagnetic field, $\mid\xi_{eR}\mid \sim 3\times
 10^{-5}$, hence we adopted $\xi_{eR}\sim 10^{-4}$ as the reference
 value in Eqs.(\ref{arbitrary}),(\ref{arbitrary2}) above.
For a topologically non-trivial 3D-hypermagnetic field configuration
with linkage (Gauss) number $n\gg 1$ for which the pseudoscalar ${\bf
  B_Y}\cdot(\nabla\times {\bf B_Y})\sim n$ one can expect the right
electron asymmetry at the level $\xi_{eR}\sim 10^{-4}$, which will be
used below as an estimated reference value with respect to which we
choose to normalize the right electron asymmetry $\xi_{eR}$.

\section{Hypermagnetic helicity evolution }
\label{sec:magn-helic-evol}
\vskip0.1cm

Let us now turn to the evolution of the hypermagnetic (magnetic)
helicity through various stages in the evolution of the universe, as
illustrated in Fig.~\ref{1099}.

\subsection{Hypermagnetic helicity evolution through the electroweak
  phase transition}
\label{sec:hyperm-helic-through}

Let us note that for Higgs masses $m_H> 80~GeV$ the electroweak phase
transition cannot be first order in the minimal standard electroweak
theory, so that a smooth cross-over between symmetric and broken
phases is more likely~\cite{Kajantie:1996mn}, given the experimental
lower bound on Higgs masses and the recent hints from the
LHC~\cite{125-higgs}. 

However, in the presence of strong hypermagnetic fields in the
primordial plasma the dynamics of the phase transition changes in
analogy with the superconductivity in the presence of magnetic fields:
the second order phase transition may become first
order~\cite{Giovannini:1997eg}. We rely here on such scenario assuming
the presence of strong hypermagnetic fields for which first order EWPT
becomes allowed in the mass region $80~GeV<m_H<160~GeV$ (see Fig. 8 in
\cite{Giovannini:1997eg}) indicated by current LHC data.

We first consider what happens with the hypermagnetic helicity
passing-through the electroweak phase transition. For that let us
mention results from paper ~\cite{Akhmet'ev:2010ba} where the flow of
the hypermagnetic helicity in the embryo of the new (broken) phase was
considered. If a single bubble of broken phase appears at $T=T_{EW}$
growing with constant velocity, $R(t)=v(t-t_{EW})$~\footnote{Here time
  is fixed at $T_{EW}$, $(t-t_{EW})/t_{EW}\ll 1$, $v=0.1-1$
  accordingly~\cite{Kibble:1995aa,PhysRevD.57.664}.}, then a value of
the flow of hypermagnetic helicity density through the bubble surface
is determined by the surface integral in Eq.~(\ref{helicity}) which we
neglected above for the symmetric phase with boundary at infinity. The
result (Eq. (17) in ~\cite{Akhmet'ev:2010ba}) shows that the value of
the hypermagnetic helicity density penetrating the surface of a single
bubble,
\begin{equation}\label{value}
\frac{h_Y(t)}{G^2cm}= \frac{5\times 10^{-3}n}{d(cm)}\left(\frac{B_Y(t_{EW})}{1~G}\right)^2\left(\frac{t-t_{EW}}{t_{EW}}\right)^2,
\end{equation}
is also large. Here the integer $n=\mp 1, \mp 2,...$ denotes the number of pairs
of linked hypermagnetic field loops (or knot number) for the non-trivial
3D-configuration, as in Eq. (7) in
Ref.~\cite{Akhmet'ev:2010ba}. Note that $n$ is the pseudoscalar
entering the Gauss integral for magnetic helicity $$H(t)=\int
d^3xh({\bf x}, t)=n\Phi_1\Phi_2$$ which changes the sign, $n\to - n$,
after one of the overlapping oriented loops in a pair of magnetic closed tubes
changes the direction.

In order to avoid screening of the hyperelectric field ${\bf E }_Y$
and the time component $Y_0$ over the surface of the bubble the
thickness $d$ of the domain wall separating the two phases should be
less than the Debye radius, $d<r_D=\sqrt{3T_{EW}/4\pi e^2n_e}\sim
10/T_{EW}$, which allows us to estimate the factor $d^{-1}$ in
eq.~(\ref{value}) as $[d(cm)]^{-1} > 10^{15}/2$. Then substituting the
value of the hypermagnetic field $B_Y(t_{EW})$ estimated in in the
leptogenesis scenario as $B_Y(t_{EW})\sim 5\times 10^{17}~ G$
\cite{Semikoz:2009ye,Semikoz:2011tm} , one gets from Eq. (\ref{value})
the helicity density $$h /G^2cm = 6.25\times
10^{47}n[(t-t_{EW})/t_{EW}]^2.$$ Such huge value is estimated at the
moment of the growth of a bubble of the new phase, e.g. for $R(t)/l_H<
[(t-t_{EW}/t_{EW})]\sim 10^{-7}$~\footnote{Such bubble size is
  relevant before percolation (collision and following junction) of
  the two bubbles, see Eq. (21) in
    Ref.~\cite{PhysRevD.57.664}.}. Taking into account the subsequent
conservation of the net global helicity, summed over different
protogalactic scales, one finds values that are much larger than the
helicity density associated to galactic magnetic field
strengths, $$h_{gal}\sim 10^{11}~G^2cm.$$

During the electroweak phase transition the hypermagnetic helicity
density converted from the symmetric phase to the Maxwellian one is
redistributed within a bubble over its volume in correspondence with
the ratio of volumes
$$V_{surface}/V_{ball}=3d/R(t).$$  This is because we calculated the
helicity flux density in eq.~(\ref{value}) only within a thin
spherical layer $V_{surface}=4\pi R^2(t)d$. Then assuming that all
bubbles are tight-fitting each other within the horizon size
$l_H(t_{EW})=1.44~ cm$ or the mean magnetic helicity density coincides
with that within one bubble, and using the ratio $3d/R(t)$ we obtain
from (\ref{value}) at $T\simeq T_{EW}$ the magnetic helicity density
in the new (broken) phase
\begin{eqnarray}\label{value2}
&&\frac{h(x\sim 1)}{G^2~cm}=\frac{1.5\times 10^{-2}n}{(l_H(t_{EW})/1~cm)}
\left(\frac{B_Y(t_{EW})}{1~G}\right)^2\times\nonumber\\&&\times\left(\frac{R(t)}{l_H(t_{EW})}\right)=
\frac{1.5\times 10^{-9}n}{1.44}\left(\frac{B_0^Y}{1~ G}\right)^2\times\nonumber\\&
&\times\exp\left[\frac{166(\kappa -1)}{\kappa^2}\int_{1}^{\infty}\frac{{\rm d}x}{x^{2}}\left(\frac{\xi_{eR}(x)}{0.0001}\right)^2\right],
\end{eqnarray}
where we have substituted $R/l_H=10^{-7}$ as an estimate of the
beginning of percolation (junction) of bubbles (see Eq. (21) in
Ref.~\cite{PhysRevD.57.664}).

We now turn to a brief discussion of the bounds on the topological
linkage number  $n=\pm 1, \pm 2,...$ ($|n| >1$) in eq.~(\ref{value2}).

Starting from Gauss formula,
$$H=n\Phi^2=nB^2\pi^2\Lambda^4~,$$ and
substituting the corresponding helicity density $h=3H/4\pi R^3$, and
using the maximum helicity density $h_{\rm max}=B^2\Lambda$ one can
find a bound on ``n'' from the requirement that $h<h_{\rm max}$:
\begin{equation}\label{bound2}
n<\frac{4}{3\pi}\left(\frac{R}{\Lambda}\right)^3=\frac{4\times 10^6}{\kappa^3}\left(\frac{\xi_{eR}(T_{EW})}{0.0001}\right)^3,
\end{equation}
where we have substituted $R=10^{-7}l_H(T_{EW})$ and
$\Lambda=\kappa\eta_Y/\alpha_Y=3.3\times
10^6\kappa/[T_{EW}(\xi_{eR}/0.0001)]$.

On the other hand, from the same bound $h<h_{\rm max}$ using Eq.
(\ref{value2}) and cancelling $B^2(t_{EW})$ one finds:
$$
\frac{h(x=1)}{B^2\Lambda}= \frac{1.5\times 10^{-9}n(\xi_{eR}(T_{EW})/0.0001)}{2.88\times 3.3\times 10^6\kappa \times 10^{-16}}< 1,
$$
or
\begin{equation}\label{bound3}
\kappa> 1.6 n\left(\frac{\xi_{eR}}{0.0001}\right).
\end{equation}

Combining (\ref{bound2}) and (\ref{bound3}) from the chain of
inequalities we get,
$$
n<\frac{4\times 10^{6}}{\kappa^3}\left(\frac{\xi_{eR}(T_{EW})}{0.0001}\right)^3<\frac{4\times 10^{6}}{(1.6)^3n^3},
$$
hence we find an upper bound on the linkage number
\begin{equation}\label{bound4}
 \mid n\mid< 33~.
\end{equation}
Note that it does not depend on a right electron chemical potential
nor on the bubble size before percolation, chosen in
Eq.~(\ref{bound2}) as
$R=10^{-7}l_H(T_{EW})$~\cite{Kibble:1995aa}. Indeed, the scale
$\Lambda\sim \kappa$ is proportional to the bubble size $R$,
$\Lambda\sim R$, as seen from Eq.~(\ref{bound3}) where $\kappa\sim R$
through the helicity $h(x=1)$ in Eq. (\ref{value2}).

\subsection{Hypermagnetic helicity evolution before 
the electroweak phase  transition}
\label{sec:hyperm-helic-evol-before}

For an arbitrary scale $\Lambda=\kappa\eta_Y/\alpha_Y$, $\kappa>1$, we
obtain from eq.~(\ref{arbitrary2}) the hypermagnetic helicity in the
unbroken phase ($x\geq 1$) as
\begin{eqnarray}\label{hyperhel}
&&\frac{h_Y(x)}{G^2~cm}=\nonumber\\
&&-0.88\times 10^{-8}\left(\frac{B_0^Y}{1~
G}\right)^2\int_x^{\infty} \frac{{\rm
d}x^{'}}{x^{'3}}\left(\frac{\xi_{eR}(x^{'})}{0.0001}\right)\times\nonumber\\&&\times
\exp \left[\frac{166(\kappa
-1)}{\kappa^2}\int_{x^{'}}^{\infty}\frac{{\rm d}x^{''}}{x^{''2}}
\left(\frac{\xi_{eR}(x^{''})}{0.0001}\right)^2\right],
\end{eqnarray}
where we put $x_0=\infty$, substituted $\alpha_Y(t)$ from
eq.~(\ref{alpha}), and used the expansion time
$t=(M_{Pl}/1,66\sqrt{g^*})/2T^2=M_0/2T^2$, omitting the initial
helicity value $H_Y(t_0)$.

\begin{figure}
  \includegraphics[width=7cm]{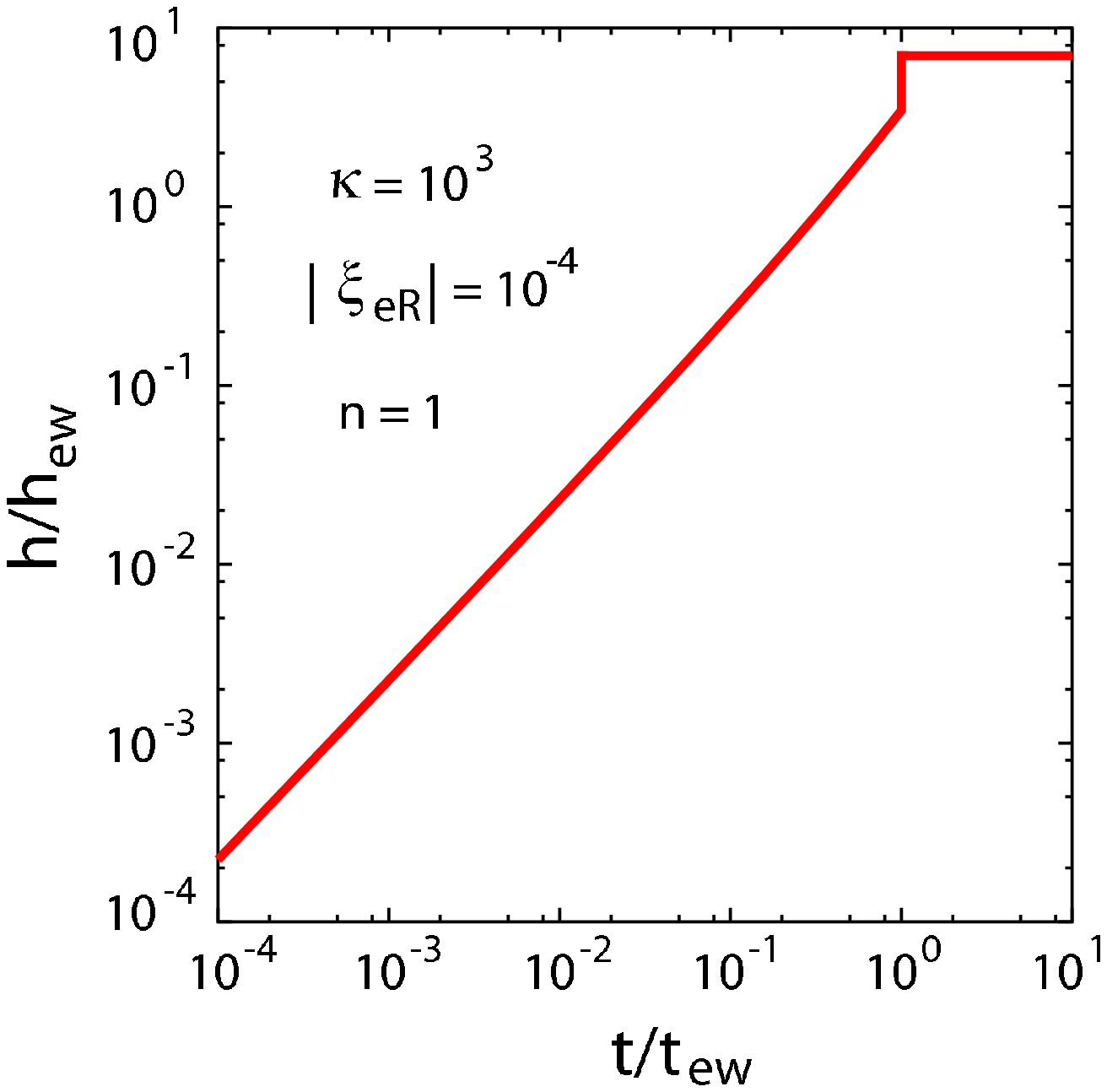} \\
  \includegraphics[width=7cm]{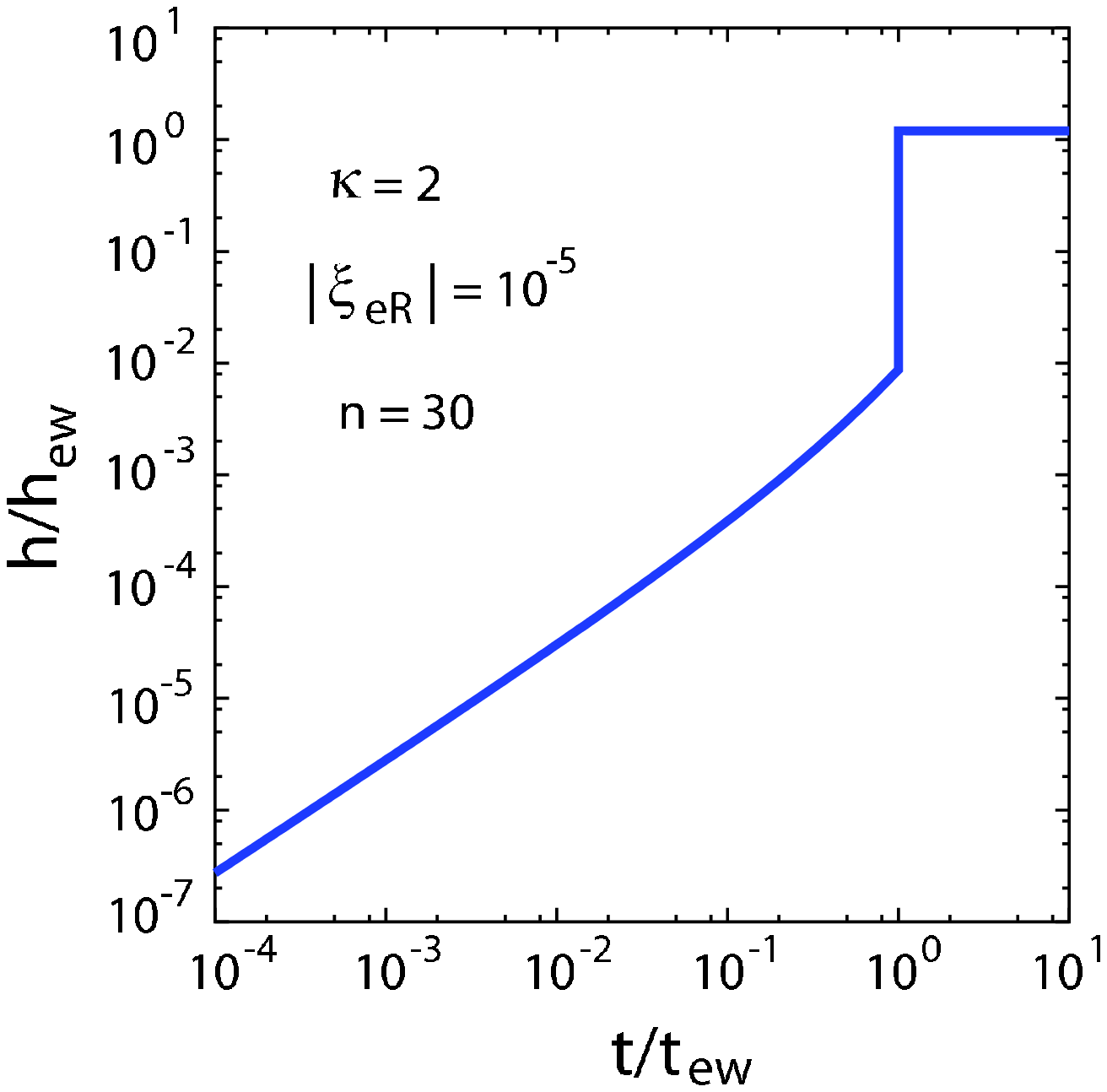}\\
  \includegraphics[width=7cm]{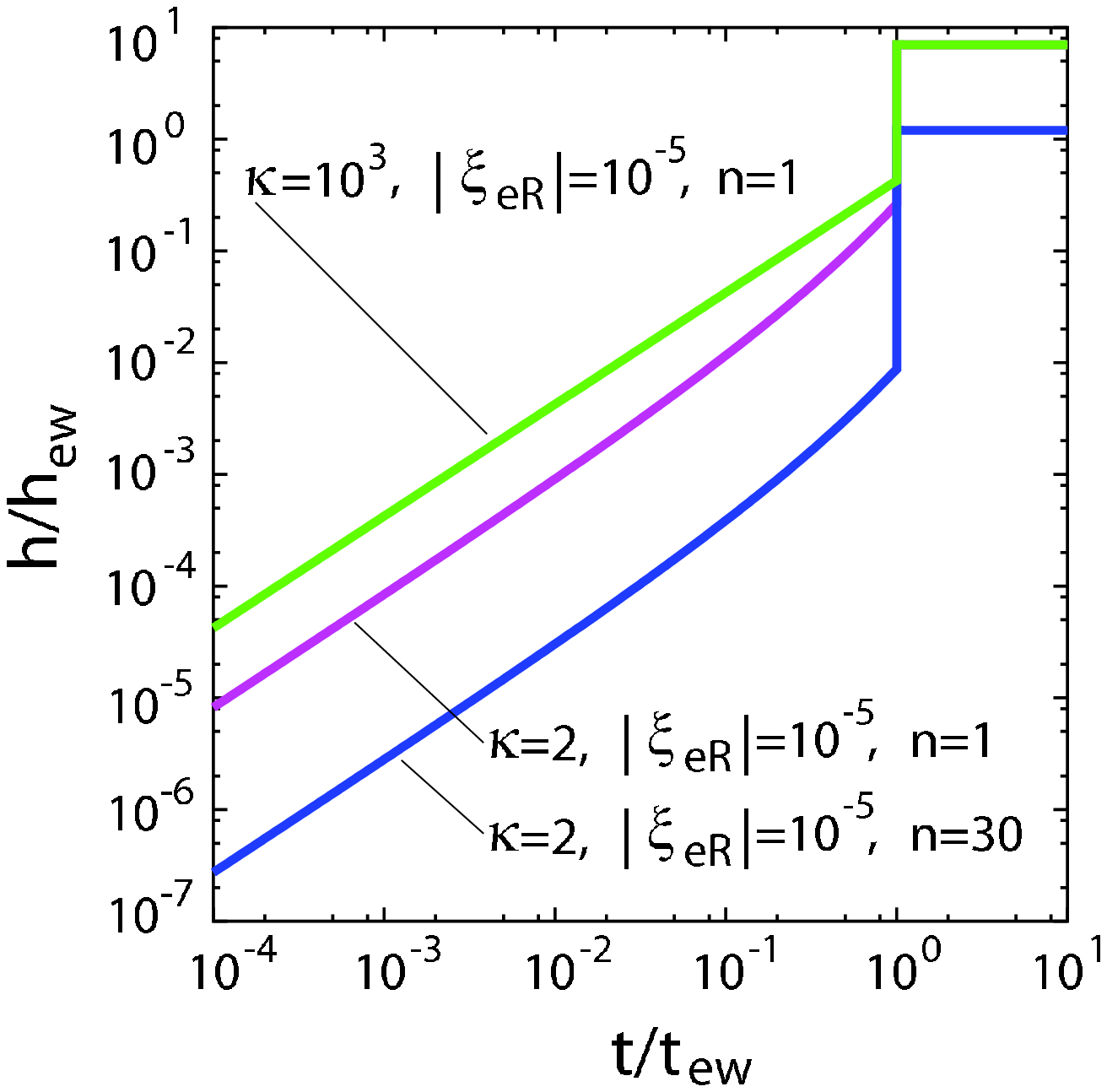}
  \caption[]{\label{1099} Magnetic helicity evolution in the EWPT epoch
    for various parameter choices: top: $a=0.166$, $\mid\beta\mid = 1$,
    $n=1$; middle: $a=41.375$ (i.e. $\kappa =2$), $\mid\beta\mid = 0.1$,
    $n=30$; bottom: red line $a=41.375$ , $\mid\beta\mid = 0.1$, $n=1$,
    blue $a=41.375$, $\mid\beta\mid = 0.1$, $n=30$, green: $a=0.166$
    (i.e. $\kappa = 10^3$), $\mid\beta\mid = 0.1$, $n=1$; for $t>t_{EW}$
    red and green lines are identical. }
  \end{figure}

  Dividing the helicity density (\ref{hyperhel}) by its value at
  $T_{EW}$ (\ref{value2}) we get the ratio valid at $x\geq 1$:
  \begin{eqnarray}\label{ratio1}
  &&\frac{h_Y(x)}{h(x=1)}=- \frac{8.5}{n}\int_x^{\infty}\frac{dx^{'}}{x^{'3}}\left(
  \frac{\xi_{eR}(x^{'})}{0.0001}\right)\times\nonumber\\&&\times\exp \left[\frac{166
  (\kappa -1)}{\kappa^2}\int_{x^{'}}^{1}\left(\frac{\xi_{eR}(x^{''})}{0.0001}\right)^2\frac{{\rm d}x^{''}}{x^{''2}}\right],
  \end{eqnarray}
  that for a constant value of the right electron asymmetry
  $\xi_{eR}/0.0001=\beta={\rm const}$ equals to
  \begin{eqnarray}\label{ratio2}
  &&\frac{h_Y(x)}{h(x=1)}= \frac{8.5\mid\beta\mid
  }{n}\left(\frac{1}{a\beta^2}\right)\Bigl[\frac{\exp[-a\beta^2(1-1/x)]}{x} -\nonumber\\&&-
  \frac{\exp [-a\beta^2(1-1/x)] - \exp (-a\beta^2)}{a\beta^2}\Bigr],
  \end{eqnarray}
  where $a=166(\kappa - 1)/\kappa^2$ and we took into account the
  negative sign of the right electron asymmetry, $\beta<0$, as seen,
  e.g., from Eq. (\ref{solution}). For a small parameter $a\beta^2\ll 1$,
  or $\kappa\gg 166\beta^2$ that is allowed for larger bubbles $R\gg
  10^{-7}l_H$ one gets from (\ref{ratio2}) (using  relation
  $t/t_{EW}=(T_{EW}/T)^2=x^{-2}$) the temporal dependence
  \begin{eqnarray}\label{ratio3}
  &&\frac{h_Y(t<t_{EW})}{h(t_{EW})}=\frac{4.25\mid \beta\mid
  }{n}\Bigl[(1- a\beta^2)\frac{t}{t_{EW}} - \nonumber\\&&-\frac{a\beta^2}{3}\left(\frac{t}{t_{EW}}\right)^{3/2}
  +O ((a\beta^2)^2)\Bigr] ,
  \end{eqnarray}
  which is almost linear in the region $t<t_{EW}$. Thus, we see from
  (\ref{ratio3}) that like for EWPT of the first order there is the {\it
    jump} of helicity density (here for large scales $\kappa\gg
  166\beta^2$):
  \begin{equation}\label{jump}
    \frac{h_Y(t\lsim t_{EW})}{h(t_{EW})}=\frac{4.25\mid \beta\mid}{n},
  \end{equation}
  where the linkage number $\mid n\mid>1$ has the upper bound given by
  Eq. (\ref{bound4}). Note that the straight lines in Fig.~\ref{1099}
  shown for some particular parameters: $a$ given by $\kappa=2,~10^3$;
  $\mid\beta\mid =0.1,~ 1$; $n=1, ~30$ correspond at $t<t_{EW}$ to the
  limiting case presented in Eq.~(\ref{ratio3}).

  Let us stress that our assumption of a constant right electron
  asymmetry $\xi_{eR}/0.0001=\beta={\rm const}$ is provided by the
  adiabatic condition $\partial_t\xi_{eR}\approx 0$ which we tacitly
  assume.

  \section{Discussion}

  The lepton asymmetry plays a crucial role in the amplification of
  hypermagnetic fields, which become helical and supply magnetic
  helicity to the cosmological and subsequently to the galactic
  magnetic fields.  The lepton asymmetry itself evolves due to the
  Abelian anomaly for the right electron current given by
  Eq. (\ref{Abelian}) in the unbroken phase of the Standard Model. It
  starts from an initial $\xi_{eR}(t_0)$ generated by an unspecified
  mechanism and then rises driven by $\alpha_Y$-helicity parameter
  (\ref{alpha}) arising from the Chern-Simons term (\ref{CS}). One
  should note that the absence of the Chern-Simons term in the broken
  phase~\cite{Laine:1999zi} does not mean that parity violation in
  electroweak interactions disappears. Indeed the polarization effect
  leading to the helicity $\alpha$ parameter governing Maxwellian
  field evolution exists due to paramagnetism of fermions populating
  the main Landau level~\cite{Semikoz:2003qt}.

  The magnetic helicity parameter $\alpha$ which governs the evolution
  of the magnetic helicity after the electroweak phase transition is
  shown in Fig.~\ref{1099} by short horizontal lines for
  $t>t_{EW}$. This change of the magnetic helicity density profile
  $h(t)$ is explained by a negligible value of the helicity parameter
  $\alpha$ for weak interactions in broken phase at $T\ll T_{EW}$,
  $\alpha\sim G_F$ ~\cite{Semikoz:2004rr,Semikoz:2003qt}, comparing
  with $\alpha_Y$ given by Eq. (\ref{alpha}) for symmetric phase,
  $\alpha\ll \alpha_Y$ .

  The jump of helicity density at $t=t_{EW}$ is the topological effect
  of a difference between the volume helicity density entering the
  first line in Eq. (\ref{helicity}) and the surface helicity term in
  the same equation.  While the volume term gives the smooth function
  (\ref{hyperhel}) in the numerator of the ratio in Eq. (\ref{ratio1})
  the surface term leads to the helicity density at EWPT $\sim n$
  given by (\ref{value2}) in the denominator of Eq. (\ref{ratio1}).

  Note also that, following standard practice, we have neglected
  turbulence due to a non-zero plasma vorticity arising e.~g.  through
  bubble collisions during the electroweak phase transition. Such
  simplification is justified in the treatment of hypermagnetic helicity
  since the fluid velocity ${\bf V}({\bf x},t)$ does not contribute to
  helicity evolution.
  As a result here we have confined our attention only to the
  $\alpha^2$-dynamo mechanism, avoiding $\alpha\Omega$ -dynamo
  scenario.

  In summary, we have found that the magnetic helicity which becomes an
  inviscid invariant after the EWPT, varies dramatically before this
  phase transition. Due to this, the cosmological magnetic field becomes
  helical before the phase transition and remains helical after it, at
  least within the range of applicability of our causal scenario. It
  means that the seed magnetic field for the galactic dynamo if provided
  by a primordial cosmological magnetic field should be substantially
  helical. This seed magnetic helicity must be taken into account in
  scenarios of galactic magnetic field evolution with a cosmological
  seed. In particular, the intergalactic magnetic field suggested in
  \cite{Neronov:2009gh} is expected to be substantially helical.

  We thank M.Shaposhnikov for fruitful discussions. V.B.S and
  D.D.S. are grateful to the AHEP group of IFIC for hospitality.  This
  work was supported by the Spanish MEC under grants FPA2011-22975 and
  MULTIDARK CSD2009-00064 (Consolider-Ingenio 2010 Programme), by
  Prometeo/2009/091 (Generalitat Valenciana), by the EU ITN UNILHC
  PITN-GA-2009-237920.

  \bibliographystyle{h-physrev4}

\end{document}